# Strategic management analysis: from data to strategy diagram by LLM

Richard Brath*  
Uncharted Software, Inc.

David Jonker*  
Uncharted Software, Inc.

Adam Bradley*  
Uncharted Software, Inc.

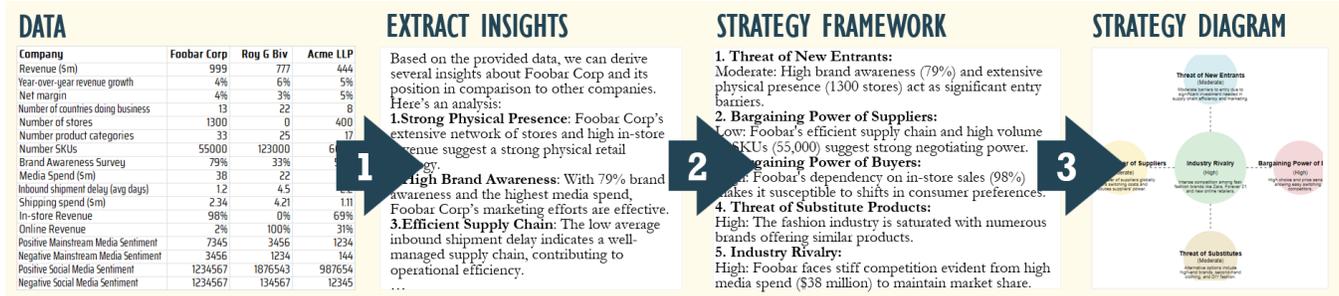

Figure 1: From insight generation to diagram by LLM: 1. The LLM generates insights from data. 2. The LLM organizes insights by a strategy management analysis framework, e.g. Porter's Five Forces. 3. The LLM generates the corresponding strategy management diagram.

## ABSTRACT

Strategy management analyses are created by business consultants with common analysis frameworks (i.e. comparative analyses) and associated diagrams. We show these can be largely constructed using LLMs, starting with the extraction of insights from data, organization of those insights according to a strategy management framework, and then depiction in the typical strategy management diagram for that framework (static textual visualizations). We discuss caveats and future directions to generalize for broader uses.

## 1 INTRODUCTION

The authors have worked on three projects involving rule-based, data-driven analytical insights for an insurance company, a mutual fund company, and a financial data company. In these projects, an insight synthesizes multiple data points into a high-level observation, such as trends over time, comparisons to peers or benchmarks, analytic models, or a document summary point. For example, an "earnings surprise" is the difference between a company's actual earnings and expected earnings estimates from analysts. No difference is not an insight, but a significant difference is (e.g., compared to prior surprises for that company or peer companies).

This conceptualization aligns with Ding et al's QuickInsights: an approach to automatically find interesting patterns from multi-dimensional datasets [1]; and is similar to John Schwabish's *takeaway message*: "telling the reader what conclusions can be drawn from the data" [2, p.36].

In large scale information environments, such as the above projects, there are many challenges with insights, such as 1) there are potentially hundreds of insights, emerging from many datasets and many dashboards; 2) organization of these insights, into groups or ranking, relevant to the specific user objective; 3) presentation of these insights.

**Strategy management** is part of management science in business schools and practiced by management consulting organizations.

"*By providing a* **framework** *that allows us to examine the* **factors** *that influence a strategic situation and organize relevant information, strategy analysis places us in a superior position to a manager who relies exclusively on experience and intuition.*" – Robert Grant, *Contemporary Strategy Analysis*, 2018 [3].

The **frameworks** used in strategy management are various forms of comparative analyses of these derived factors. There are dozens of frameworks, for different strategic analyses. A simple, popular framework is the SWOT matrix—a technique used to identify strengths, weaknesses, opportunities, and threats. It is used to identify the internal and external factors that affect the achievement of project objectives (e.g. Figure 2).

| PepsiCo SWOT Analysis | |
|---|---|
| **Strengths** | **Weaknesses** |
| • Diverse product portfolio with over 23 billion-dollar brands | • Heavy reliance on North American market (58% of revenue) |
| • Strong global presence, operating in more than 200 countries | • Negative health perceptions of core carbonated soft drink products |
| • Robust financial performance with $79.47 billion revenue in 2023 | • Exposure to commodity price fluctuations affecting profit margins |
| • Strong brand recognition with Pepsi ranking among top global brands | • Lower market share in some beverage categories compared to Coca-Cola |
| **Opportunities** | **Threats** |
| • Expansion in emerging markets with growing middle-class populations | • Increasing health consciousness leading to reduced soda consumption globally |
| • Investment in healthier product lines to meet changing consumer preferences | • Stringent government regulations on sugary drinks and snack foods |
| • Leveraging e-commerce for direct-to-consumer sales growth opportunities | • Intense competition from both global and local brands in key markets |
| • Adoption of sustainable practices to appeal to environmentally conscious consumers | • Environmental concerns and regulations regarding packaging and water usage |

Figure 2: Example SWOT analysis.

These **factors** of a strategy analysis are data-driven analytical statements highly similar to the previously defined *insights*. An example human-authored factor in a business analysis is: "UPS is more labor intensive with a much higher ratio of employee costs to sales than FedEx." (example from [3, capsule 2.3]). The framework constrains the relevant insights and the human-author organizes the most relevant ones. The results facilitate comparison across insights and aid decision-making trade-offs.

* e-mail: {rbrath, djonker, abradley}@uncharted.software

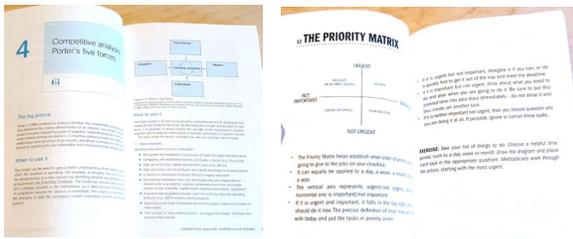

Figure 3: Samples from books itemizing strategy management frameworks with corresponding diagrams [4,5].

Furthermore, most strategy management frameworks have corresponding **strategy management diagrams** (e.g. Fig. 3). These visualizations group insight statements into a diagram. Each framework has a different visualization, which can be found in strategy management textbooks (e.g. [3]), or more briefly itemized and described in some publications (e.g. [4-5] in Fig. 2).

The combination of the analytical strategy management frameworks together with corresponding diagrams is useful for analysis and also collaboration: *"Management models are not only tools for professionals and managers but essentially they are a way of **communicating**. Management models bridge differences in abstraction and provide comprehensiveness."* – Dr. Theo Camps, Chairman, Berenschot Group [4].

Note that there is not a definitive list of strategy management frameworks, nor their diagrams. Terminology may vary between publications. Variations in diagrams may exist.

Our contribution is what we have learned in creating strategy management diagrams directly from data, using LLMs in each of three steps: 1) extract insights from various data; 2) organize insights via strategy management frameworks; and 3) draw the resulting strategy management diagram.

## 2 MOTIVATION

We have many interests in strategy management diagrams:

1) We have experience using business strategy diagrams to justify visualization projects. We now seek tools to enhance the analysis, synthesis, and generation of these diagrams.

2) Statements used in strategy analyses are derived from data and are presented textually. Organizing and presenting many textual facts could significantly benefit future text, LLM, and qualitative visualizations.

3) Strategy management diagrams are rarely discussed in the visualization community or STEM education. However, they are prevalent in management science, MBA programs, and large organizations. Re-using already well-understood visualizations increase adoption of visualization for new kinds of analyses.

4) Some users describe "dashboard hell," where an excess of charts overwhelms them, making it hard to find what they need. By considering insights as takeaway messages from charts, which can be organized into strategy management diagrams, it may be feasible to create more effective dashboard workflows.

## 3 BACKGROUND

LLMs and AI have been used in the visualization community to generate visualizations; captions, annotations and stories; or both.

For example, Yilin et al use generative AI, given data to suggest relevant visualizations [6]. Dibia's LIDA tool generates visualizations and infographics [7]. Yuan et al created ChartGPT to generate multiple visualizations in response to natural language questions [8]. Pingchuan et al generate insights that complement visualizations [9].

Regarding insights, titles, captions and annotations, Schwabish and others have noted the importance of good titles and annotations, e.g. [1]. Data facts are similar to insights [10], but include simple descriptive statistics as opposed to "takeaway messages".

DataShot extracts data facts and generates Fact Sheets (i.e. a collection of infographics, each communicating a fact [11]); and Calliope has a similar approach, generating data stories [12]. Both use similar steps of fact extraction, composition, and visual synthesis. Fact sheets and data stories are dissimilar from strategy management frameworks and diagrams, in that composition of facts is based on data attributes (as opposed to a strategy framework, semantically organizing insights into specific analytic themes) and the visuals are infographics or simple annotated charts (as opposed to strategy diagrams organizing statements by the analytic theme and typically simple graphics such as grids or connecting arrows).

There has been much research in generating insights to use with captions or annotations in visualizations. Pre-LLM captions go back to Mittal et al in 1998 [13], to much more recent systems, e.g. [14-16]; and recent research to generate captions via LLMs, e.g. [17]. ChartStory automatically sequences charts, extracts insights assembled into stories, but requires a singular dataset and the insights are tied to each chart, not a comparison of disparate insights [18]. Furthermore, using significant amounts of text within quantitative visualizations has also been investigated, e.g. Stokes et al [19-20], generally to positive effect.

As such, none of the existing approaches leverage strategy management approaches which may aid in the collect, organize and presentation insights.

## 4 METHOD AND EXAMPLES

We investigate the generation of strategy diagrams in three steps as we believe there are different kinds of potential LLM failures at each step. As outlined in Figure 1, ① we start with LLM generation of insights, from either provided tabular data or LLM knowledge (Section 4.1). ② Then we organize insights with strategy management analyses (4.2). ③ These organized insights are then represented in the traditional management diagrams associated with each of these analyses (4.3).

For ② and ③, we use four strategy management frameworks: SWOT, Porter's Five Forces, virtuous circle and Value Discipline – chosen because of the assumed increasing complexity of each:

**SWOT** is highly common, as it is used for strategy management as well as project management. It is typically visualized as a 2x2 grid with text bullets in each quadrant: graphically it is very simple.

**Porter's Five Forces** groups factors (insights) into five categories. Visually, it is typically arranged with boxes of text insights; one box central, with a box on each side of the central box, connected with arrows (Figure 2 left).

**Virtuous circle** is less common, but the authors have attended multiple meetings with CEO's pronouncing their virtuous circle business strategy. The typical depiction is a set of text boxes arranged in a circle connected with arrows.

The **Value Discipline** of Treacy and Wiersema is less common and requires reducing all insights into three predefined disciplines: operational excellence, product leadership and customer intimacy. These categories are thematically narrower than "strength" or "opportunity", thus possibly more difficult for the LLM. The traditional depiction is a radar plot, thus requiring the conversion of multiple textual insights into quantitative values, thus possibly difficult for an LLM ("LLMs don't do math" [21]).

We use multiple LLMs—ChatGPT 3.5, 4, Cohere, Gemeni, Claude and Microsoft Co-pilot in PowerPoint—as we have been investigating these for various non-research projects and have some familiarity. We use the same simple prompt across all LLMs (e.g. *Do a SWOT analysis of Walmart*), as opposed to a more verbose prompt with step-by-step instructions, as the more tailored prompt may be well suited to one LLM but perform poorly for others.

## 4.1 LLM insight generation

We started with insight generation via LLMs, using either preexisting A) LLM knowledge gained from training data; or B) direct input of quantitative data. For the initial task, we focused on financial data as we have many decades of expertise working with financial analysts, and have previously worked on insight systems in financial services with hundreds of insights.

**A. LLM insights from prior training data**. LLMs can reuse the knowledge of their training data. We use the prompt *"Based on your training data knowledge, describe the recent trend in the income statement from <company>."* We use *income statements*, as they are highly standardized accounting tables embedded in quarterly documents filed with federal regulators, easily downloadable and highly likely to have been used in the training of massive LLMs; and thus likely for LLMs to be responsive to.

The *trend* keyword restricts the LLM to use temporal data found within the income statement, thereby not requiring the LLM to compare and contrast against peer companies and benchmarks, which makes validation more difficult. The initial clause is used overcome the LLM refusal to answer based on lack of recent data.

An example answer from ChatGPT 3.5 is:

*As of my last update in July 2023, Walmart's income statement has reflected several key trends over recent periods:*
- *Revenue Growth: Walmart has generally experienced steady revenue growth, driven by a combination of factors including its strong e-commerce presence, expansion in international markets, and robust performance in its grocery segment.*
- *E-Commerce Expansion: E-commerce sales have been a significant growth area for Walmart, with increased investments in online infrastructure and partnerships (such as with Shopify). This segment has seen strong year-over-year growth, contributing significantly to overall revenue.*

… and five more insights.

The correctness of these insights can be validated by corresponding web searches as reports and analyses based 2023 Walmart performance can be found online. Their relevance as insights can be compared against the insights in the prior rule-based system. Of the seven insights generated from the income statement, five directly correspond to rule-based insights and the remaining two are related to similar rules – for example e-Commerce above is similar to a revenue segmentation rule. This analysis can be run for other companies, including less-well-known American public companies with similar results.

The derivation of these insights, however, cannot be deduced from the LLM. Rather than being derived from tabular data, they could be synthesized from various reports from stock market analysts used in ChatGPT's training data. Therefore, we consider insight generation from data provided in the prompt:

**B. LLM insights on provided timeseries data.** We gave LLMs the prompt *"Describe the trend in this timeseries data,"* followed by the previous 30 days of tab-delimited timeseries data for one stock with date, stock price and volume. This can be validated to be outside the training data of the LLM assuming the LLM correctly responds to *"When was your training data last updated?"*

The output can be compared to prior rule-based systems. For example, ChatGPT 3.5 generated three top-level insights that matched rules in our prior insight-rule-system, and it also provided a breakdown of more micro-analysis under each top-level insight, not relevant in our prior system. Cohere generated two insights aligning with our prior system; and it also identified a weekly cycle pattern (*The data also reveals a general pattern of lower closing prices towards the end of each week, with small recoveries at the start of the following week. This weekly cycle is most noticeable in the first half of April.*). Cyclical trends were not in our prior system, but, are covered as a pattern of interest in financial timeseries analyses textbooks, e.g. [21], discoverable in seasonality charts, and commercialized by companies such as Seasonax.

**C. LLM insights on provided tabular data.** A financial analyst will also find insights in other financial data such as quarterly accounting tables (e.g. the income statement), or transcripts from earnings calls, press releases, etc. This can be assessed from viewing equity analyst reports, hundreds to thousands of which are available in financial data systems. Using the income statement table from one stock, released the prior day (and outside of training data as identified by the LLM), ChatGPT 3.5 identified a half dozen insights, a few listed here:

- *Net Income: shows a steady growth from $20,081 million in June 2023 to $21,939 million by March 2024. This indicates effective tax management and profitability from continued operations, despite changes in additional income/expense items which sometimes negatively impacted earnings before tax (notably the -$854 million in March 2024).*
- *Operating Income and Earnings Before Tax (EBT): demonstrate a general upward trend, with Operating Income growing from $24,254 million in June 2023 to $27,581 million in March 2024. Earnings Before Tax mirror this positive movement, which reflects solid overall operational control and profitability growth.*
- *Cost of Revenue and Gross Profit: The Cost of Revenue varied from $16,795 million in June 2023, decreasing initially and then increasing to $18,505 million by March 2024. This indicates fluctuating production or operational costs which need to be monitored for better cost management. Gross Profit has consistently increased from $39,394 million to $43,353 million over these quarters, which implies an improvement in cost efficiency or pricing strategies, outpacing the growth in the cost of revenue.*

LLM analyses of unseen income statements suggests LLMs may be capable of direct insight generation of standardized tabular data. But can LLM-derived insights be extracted from non-standardized datasets? We gave the LLMs a non-standard synthetic tiny tabular dataset (Figure 1 left) with 14 metrics for a fictious company (Foobar Corp) and two competitors. This dataset does not follow standardized tables so the LLM cannot use preexisting knowledge of those tables. The metrics are terse descriptions of data fields that might be found in financial data systems. The fields names correspond to the metric types that might be used in business—e.g. financials, retailing, inventory, customers, e.g. Number of countries doing business, Number product categories, Brand awareness survey, Media spend ($m), In-bound shipment delays, In-store revenue, Online revenue, Positive mainstream media sentiment, Negative mainstream media sentiment, Positive social media sentiment, Negative social media sentiment.

These metrics were created with varying complexity. For example: a) shipping delays can be simply compared between companies; b) revenue has two variations which can be combined for totals, analyzed individually, or analyzed as ratios; c) sentiment has four variations, which are most useful when compared: e.g. social positive/negative vs mainstream positive/ negative.

We then ask: *"Given the data below, what insights can you derive about Foobar Corp?"* ChatGPT 3.5 generated one data fact per metric, then synthesized these into six high-level insights:

1. **Strong Physical Presence**: *Foobar Corp's extensive network of stores and high in-store revenue suggest a strong physical*

*retail strategy. However, this also indicates an area of improvement in online sales, which are currently very low.*
2. **High Brand Awareness**: *With 79% brand awareness and the highest media spend, Foobar Corp's marketing efforts are effective. However, balancing the media spend with returns on investment would be crucial.*
3. **Efficient Supply Chain**: *The low average inbound shipment delay indicates a well-managed supply chain, contributing to operational efficiency.*
4. **Product Diversity**: *With the highest number of product categories, Foobar Corp offers a wide range of products, which could attract a diverse customer base.*
5. **Balanced Profitability**: *While the net margin is decent, it's essential to monitor and possibly improve profitability, especially in comparison to competitors like Acme LLP.*
6. **Public Perception**: *Managing and improving public sentiment, especially on social media, is important for maintaining and enhancing brand reputation.*

Gemini and Claude instead grouped the 14 metrics into eight categories: e.g. market size and growth, profitability, geographic reach, product diversification, brand and marketing, supply chain efficiency, sales channels, and customer sentiment; and then providing one to three synthesized insights per each group.

We can manually inspect the resulting statements against the source data to confirm their validity, for example, Foobar *does* have higher in-store revenue than competitors and low online sales.

## 4.2 Organizing insights via management frameworks

Given a number of insights, as in the examples above, there are many possible ways to organize these. It is feasible to present the most impactful insight(s) as a caption, title or annotations within a visualization, or as a narrative associated with a visualization—as discussed in many visualization research papers. But, insights may not align with charts; nor, insights may originate from tables (e.g. income statements) or text not typically visualized.

As discussed in the introduction, *strategy management frameworks*, organize insights (factors) for comparative analysis, according to the criteria of the particular framework. A framework organizes insights, and, also reduces the insights to those insights which are the most relevant as supporting evidence for each theme within the framework. For example, if an insight does not align or is only weakly aligned with any of the themes of strength, weakness, opportunity and threat in a SWOT framework, it would not be included (or listed below the stronger insights.)

The transformation of insights from simple "data fact observations" into insights of "interesting patterns" exists in the analytic pipelines of QuickInsights, DataShot, Calliope, [1,11,12] and our earlier rule-based work. In strategic management frameworks, the framework is the forcing function to reduce a large number of insights into a short list of most relevant insights.

**A. LLM use of strategy frameworks on training data.** As before, we start with using the LLM to organize insights for a management framework directly from prior training data. This allows us to assess LLM familiarity and capabilities with strategy frameworks. LLMs have learned a broad variety of management frameworks—we tried SWOT, BCG Matrix, Porter's Five Forces, virtuous circle, Value Discipline of Treacy and Wiersema, and Kraljic's Purchasing Model on various large companies such as Walmart, McDonalds, fashion retailers, tech companies and other entities.

For example, a SWOT analysis for Walmart via ChatGPT3.5 generates five insights per each category of SWOT, the first two of each category listed here:

*Strengths*:
1. *Strong global presence with a vast network of stores.*
2. *Wide range of products offered, including groceries, electronics, apparel, and household goods.*

*Weaknesses*:
1. *Dependence on physical stores, facing challenges from the shift to online shopping.*
2. *Negative publicity regarding labor practices and employee relations.*

*Opportunities*:
1. *Growth in e-commerce presents opportunities for Walmart to expand its online presence.*
2. *Expansion into emerging markets, particularly in Asia and Africa.*

*Threats*:
1. *Intense competition from online retailers like Amazon and brick-and-mortar competitors like Target and Costco.*
2. *Economic downturns impacting consumer spending.*

As these insights from LLM training data are no longer connected to their training data, it is difficult to validate their responses intrinsically using the LLM. Therefore, we generated four management analyses (SWOT, Porter, BCG Matrix and Value Discipline) via ChatGPT3.5 for some of our large Fortune 500 customers, and then asked their senior management (who often work with management consultants) to comment on these LLM generated analyses of their firms. We received answers: *"Yes, that sums up common sentiment, but it misses recent work we've done in the last year,"* and *"This is light - but it is a great example of how we are on the verge of a huge shift in how human to software communication happens."* This is promising that the LLM has captured valid insights and further that these managers did not uncover any hallucinations.

**B. LLM strategy frameworks on tabular data.** We asked the LLMs to organize the analyses of Foobar Corp to various frameworks. For example, here is the first example from each SWOT category in an analysis of Foobar by ChatGPT3.5:

*Strengths*:
1. *Market Presence: Operating in 13 countries and having 1300 stores indicate a strong and widespread market presence, providing opportunities for market expansion and revenue growth.*

*Weaknesses*:
2. *Online Revenue: While Foobar Corp generates revenue from online channels, the percentage of online revenue compared to in-store revenue is not provided, which could be a weakness if online sales are not optimized.*

*Opportunities*:
1. *Market Expansion: With a global presence in 13 countries, there are opportunities for further market expansion into new regions or untapped markets.*

*Threats*:
1. *Competitive Landscape: Operating in a diverse market with 33 product categories may expose Foobar Corp to intense competition, requiring continual innovation and differentiation to maintain market share.*

Of note are some of the data handling skills or gaps. Regarding opportunities, the LLM uses its real-world of countries to assess that 13 countries represent only a subset and thus more countries are an opportunity. However, for weaknesses, it does not compare the ratio of on-line sales vs. instore sales even though the underlying data has been given.

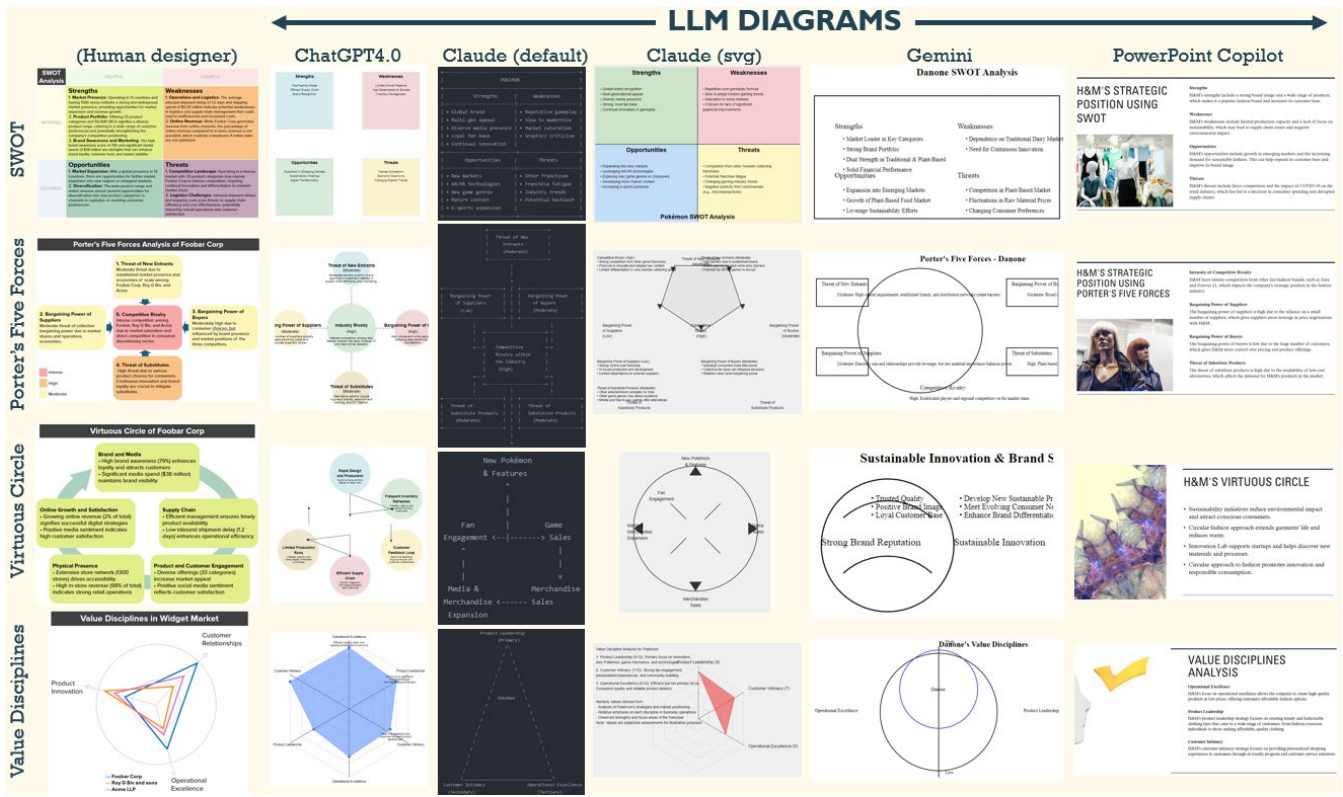

Figure 4: Four strategy management frameworks and corresponding diagrams created by a human designer and various LLMs.

Furthermore, these analyses were asked in the same session as the prior initial insight extraction on Foobar and there are variations in the resulting insights. For example, the LLM missed the ratio of online sales during the SWOT analysis whereas it used the ratio in the insight analysis (Section 4.1). As such, there are likely many cases where the LLM is generating insights from the data directly for the framework rather than organizing the prior insights.

The LLMs were expected to have difficulty with the Value Discipline framework. It requires synthesizing evidence into three pre-defined categories: customer relationships, product innovation, and operational efficiency; and also creating a numeric value for each. All the LLMs "knew" of the Value Discipline model. ChatGPT3.5 and Cohere could not determine, nor attempt to derive, the categories: *"While the dataset provides information on revenues, growth rates, margins, marketing spend, and other operational metrics, it lacks specific data related to customer relationships, product innovation, or operational efficiency."* Gemini did synthesize insights into categories but neglected to provide numeric values, despite being explicitly asked. ChatGPT4 and Claude synthesized insights into the appropriate categories but sometimes neglected to compare metrics relative to peers. The LLMs were inconsistent in which data attributes to use for which Value Discipline category. Sentiment was used by ChatGPT4 for customer intimacy, while Gemini for product leadership.

We note that the latest LLMs seem to be able to do appropriate calculations *sometimes*. For example, in our analyses, we found Cohere properly performed a percent change calculation between the open and low price in Section 4, and ChatGPT4 generated appropriate results comparing sentiment between the 4 sentiment metrics: social positive/negative vs mainstream positive/negative, for example: *Roy G Biv Product Leadership: Strong product leadership indicated by higher positive social sentiment (1,876,543 positive vs. 134,567 negative). Although the mainstream media sentiment ratio is less favorable, the considerable positive social media sentiment suggests effective innovation.* (note that social sentiment was 93% positive, which was not provided in the data; and mainstream media sentiment was 3456 positive vs 1234 negative, i.e. 73% positive. The LLM needed to pair the correct metrics, do the correct calculation, and interpret the results.)

### 4.3 Presenting insights via management diagrams

The prior two objectives were focused on generating textual content. The final objective was to generate strategy management diagrams with content. The results are shown on the right side of Figure 4. (Some columns use examples from Foobar, some columns use other popular companies).

On the vertical axis are four analyses: SWOT, Porter's Five Forces, Virtuous Circle and Value Discipline. The first column shows a human designer depicting content generated by the LLM in the prior step with the traditional diagram layout associated with that strategy framework. By "traditional diagram", we mean the diagram associated with the analysis in [3], which also corresponds to the image most frequently occurring in Google image search. E.g. for SWOT, a 2 x 2 matrix; for Value Discipline, a radar plot; and so on. Note that other layouts are possible. Our prompt explicitly asked for the "*traditional diagram, including explanatory text*," and for the Value Discipline we asked "*traditional radar plot*". For some LLMs we needed to explicitly append "*in HTML SVG*" when the LLM instead defaulted to ASCII diagrams.

Successive columns are diagrams generated by LLMs: ChatGPT4, Claude (default), Claude SVG, Gemini and Copilot in in Microsoft PowerPoint (which always defaulted to text bullets and stock photos.) The LLMs could mostly create passable grids with text (SWOT, top row), and had a variety of difficulties on the other diagrams. Further observations:

**Layout:** The LLM doesn't necessarily create the traditional layout. For Porter's five forces, only ChatGPT managed a good result, and Claude's ASCII is close. The virtuous circle was problematic, with ChatGTP coming close except for arrows and spacing, and Claude's Porter diagram a better circle than its circle diagram. For the Value Discipline, in the traditional diagram, the insights are synthesized into ordered values and represented as a radar plot with three axes. Some LLMs insisted on more axes, others struggled to get a plot.

**Text fitting:** With the exception of grids, ASCII or PowerPoint bullet lists, the LLMs struggled with fitting text into graphics. Text is overlapped, obscured by other graphics, does not fit into boxes, is chopped off by the container or group, and so on. This is an interesting area for deeper investigation: graphics libraries such SVG or D3 do not provide classes for text fitting, and further, fitting text is a optimization problem requiring adjustments to font size, box size and scaling the overall image to make everything fit. It is likely LLMs do not have this in their training data or perhaps code generation is a preferable approach to generation of text diagrams.

**Color:** The LLM either does not use color or uses arbitrary colors. For Porter's Five Forces, each top-level category has a risk indicator (e.g. intense, high, moderate), which was not used by the LLM, but the human designer used to draw attention to the riskiest category. ChatGPT did use color, but the high risk categories in ChatGPT's output are colored green in one case and red in the other, potentially leading to confusion if the diagram is skimmed.

**Text to orderable values**: The Value Discipline diagram uses a radar chart which requires transforming text statements into quantitative values. The LLMs do generate a value and thus (somewhat) able to generate a radar chart.

Given the work of other researchers with visualization generation (background section) we had incorrectly anticipated the LLM would be able to generate these diagrams with simple prompts. Further work on prompting approaches may be required.

## 5 DISCUSSION

The biggest gaps in the full sequence of data to diagram are primarily regarding: 1) Inconsistent abilities to manipulate numbers. This being addressed in LLMs through the use of plug-ins/assistants; and 2) Poor ability to generate diagrams particularly when text fitting is required. This might be addressed in a variety of ways, such as variations in prompts or output type (e.g. code generation or image generation); and/or more holistic LLMs that also incorporate multiple media types including graphics into their training data.

While prompts and next generation LLMs may solve these two issues, there are several assumptions and caveats for the future user of LLM generation of insights, strategy analyses and diagrams:

1. **Sensitivity to data structure:** In Section 4, the first timeseries dataset was organized in reverse order – the most recent date was the first row. In many years of working with timeseries data, this data is almost always organized with earlier dates occurring on earlier rows. The LLMs generated erroneous insights with rows in an unexpected order.
2. **RAG:** Using the LLM's knowledge gained from its training data may result in responses that are significantly out of date. However, approaches such as RAG and GraphRAG [23] can be combined with LLMs to generate responses effectively utilizing more recent data.
3. **Qualitative analysis via management models**: Management models are an interesting starting point for the analysis of multiple disjoint textual statements and might be extensible for more generalized qualitative analysis.
4. **Assembling qualitative results into quantitative values**: Some analyses require transforming a list of qualitative insights into a quantitative value. This seems like an area where error, subjectivity or bias may be introduced.
5. **Metrics derivation:** Many analyses require deriving appropriate metrics, known as feature engineering in machine learning. This is challenging because tabular data lacks metadata beyond column headers and values, making metric semantics hard to assess. Combining these metrics into explainable features remains unsolved. More training data from data science and management science may be needed for LLM training. Future work could compare how LLMs combine metrics under ambiguous requirements.
6. **LLM rapid improvement:** Ethan Mollick [24] claims: "Assume the LLM you are working with right now is the worst LLM you will ever work with." We generated diagrams in April, then three months later in July. Some improvements were significant: see Figure 5. Furthermore, the three steps outline in this paper can be collapsed into a single prompt: the SWOT in figure 2 was generated with: *Do a SWOT for on the company PepsiCo and create the standard 2x2 grid as HTML DIV or HTML SVG, include no more than 4 factors per cell and at least 5-10 words per each factor in each cell. Your factors can include metrics available to you as of your last update.*

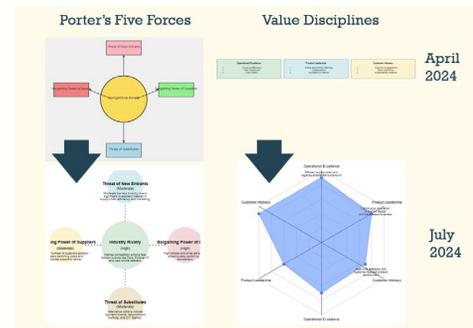

Figure 5: Improvements in generated charts (ChatGPT4).

## 6 CONCLUSIONS

We believe our investigation shows that LLM generation of strategy management diagrams are feasible with some caveats. Further, these positive results are supportive of future research of application of strategy management diagrams to a broader range of data facts and qualitative data, not strictly limited to corporate datasets. At the most extreme position, Ben Shneiderman has said *"The purpose of visualization is insight, not pictures."* [25], which is a provocation: if we can use automated techniques to extract and organize discrete insights, then the role of visualization could change significantly in analyses, and this paper is one approach for consideration of new kinds of visualizations of insights. At the other extreme, future strategy management diagrams could provide a high-level entry point to visual analytic systems, starting with 1) the user (business) questions answer through insight (strategy) diagrams, 2) augmented with word-scale graphics [26] and 3) elaborated with interactive drill-down for visualization data analysis and exploration, particularly for complex data not easily explained in simple terms (e.g. causality [27,28]).